 \documentclass[aps,prl,twocolumn,showpacs,floatfix]{revtex4}
\usepackage{graphicx}
\usepackage{dcolumn}
\usepackage{bm}
\pdfoutput=1
\begin{document}

\title{Transforming carbon nanotubes by silylation: An {\em ab initio} study }

\author{Kiseok Chang}
\affiliation{Physics and Astronomy Department,
             Michigan State University,
             East Lansing, Michigan 48824-2320, USA}

\author{Savas Berber}
\affiliation{Physics and Astronomy Department,
             Michigan State University,
             East Lansing, Michigan 48824-2320, USA}

\author{David Tom\'anek}
\affiliation{Physics and Astronomy Department,
             Michigan State University,
             East Lansing, Michigan 48824-2320, USA}

\date{\today}


\begin{abstract}
We use {\em ab initio} density functional calculations to study
the chemical functionalization of single-wall carbon nanotubes and
graphene monolayers by silyl (SiH$_3$) radicals and hydrogen. We
find that silyl radicals form strong covalent bonds with graphene
and nanotube walls, causing local structural relaxations that
enhance the $sp^3$ character of these graphitic nanostructures.
Silylation transforms all carbon nanotubes into semiconductors,
independent of their chirality. Calculated vibrational spectra
suggest that specific frequency shifts can be used as a signature
of successful silylation.
\end{abstract}




\maketitle


Carbon nanotubes\cite{TAP111} have evolved into one of the most
intensively studied one-dimensional nanostructures. They owe their
success in the nanotechnology field to a unique combination of
atomic-scale perfection, structural stability, beneficial thermal
and electrical properties\cite{TAP111}. Their usefulness as
molecular nanowires is hindered by two main limitations. First,
reproducible formation of contacts with a well-defined geometry
and electronic properties is difficult, since most materials do
not wet the surface of the inert nanotube, but rather bond to
ill-defined unterminated ends, causing large variations in the
contact resistance\cite{DT178}. Even more important is the fact
that the conductance of carbon nanotubes depends sensitively on
the precise way a graphene layer is rolled up into the tubular
structure, identified by the chiral index
$(n,m)$\cite{{Mintmire92},{RSaito92},{Oshiyama92}}. Since it is
currently impossible to produce nanotubes with a unique chiral
index, even samples with a narrow diameter range necessarily
contain a mixture of metallic and semiconducting nanotubes,
reducing their usefulness in quantum devices.

Here we propose that the recently\cite{SSWong06} achieved chemical
functionalization of carbon nanotubes by silyl groups bears
promise for the nanotube walls to form strong, well-defined bonds
to a surrounding matrix, and for the majority of nanotubes to
convert into semiconducting nanowires. Using {\em ab initio}
calculations, we demonstrate that silyl (-SiH$_3$) radicals form
strong covalent bonds with the surface of nanotubes and graphene.
Since silylation introduces a partial $sp^3$ character into the
$sp^2$ graphitic network, it opens up the fundamental band gap and
converts all carbon nanotubes to semiconductors. We show that
successful silylation of nanotubes can be verified experimentally
by monitoring specific changes in the vibrational
spectra\cite{SSWong06}.

The selection of silyl radicals as functional groups has been
motivated by the success of silane chemistry in establishing
strong, covalent bonds between polymers and inorganic
matter\cite{silane-bonding}. The rich silane
chemistry\cite{silane-book} is likely to provide ways to modify
silyl radicals bonded to the nanotube wall in order to optimize
covalent bonding of nanotube walls to many materials, including
silicon. To our knowledge, there has been no prior theoretical
study of the silyl-nanotube interaction and the effect of
silylation on the nanotube properties.

In order to determine, whether silyl radicals may covalently
attach to nanotubes and thereby modify their conductance
properties in a desirable way, we used {\em ab initio}
calculations to determine the equilibrium geometry, total energy
and electronic structure of silylated nanotubes. Our calculations
are based on the density functional theory (DFT) within the local
density approximation (LDA). We used the Perdew-Zunger
\cite{Perdew81} parameterized exchange-correlation functional, as
implemented in the \textsc{SIESTA} code
\cite{soler-tsmfaioms2002}, and a double-$\zeta$ polarized basis
localized at the atomic sites. The valence electrons were
described by norm-conserving Troullier-Martins pseudopotentials
\cite{Troullier91} in the Kleinman-Bylander factorized form
\cite{Kleinman82}.

In our calculations, we considered the products of the
dissociative adsorption of silane,
SiH$_4{\rightarrow}$SiH$_3$(ad)+H(ad), yielding silyl radicals and
hydrogen atoms chemisorbed on graphene and carbon nanotubes at
different coverages. We obtained results for carbon nanotubes of
different diameters, with chiral indices ranging from $(4,4)$ to
$(18,0)$. To describe isolated nanotubes while using periodic
boundary conditions, we arranged them on a tetragonal lattice with
a large inter-wire separation of $22$~{\AA}. Depending on the
SiH$_3$ coverage and the chiral index, we used supercells
containing up to 4 primitive unit cells in the tube
direction\cite{CNT-supercells} and sampled the Brillouin zone of
these 1D structures by at least 7~k points. We limited the range
of the localized orbitals in such a way that the energy shift
caused by their spatial confinement was no more than
50~meV\cite{SIESTA_PAO}. The charge density and the potentials
were determined on a real-space grid with a mesh cutoff energy of
$200$~Ry, which was sufficient to achieve a total energy
convergence of better than 0.1~meV/atom.

When calculating the vibrational spectrum of pristine and
silylated nanotubes, we first optimized the reference structure to
a high precision, with residual forces not exceeding
$0.04$~eV/{\AA}. The high mesh cutoff energy value guarantees that
the forces are translationally invariant. To obtain a good
estimate of the Hessian matrix in the harmonic limit, we displaced
atoms in the positive and negative direction along each axis by
the small amount of $0.01$~{\AA} and determined the forces
analytically. With these numerical precautions and limitations of
the LDA force field, we found our frequencies to be dependable
with the accuracy of down to few inverse centimeters. We also
found our LDA-value\cite{graphite-Gband-LDAGGA} for the G-band
frequency of $1630$~cm$^{-1}$ to agree much better with the
experimental value\cite{graphite-Gband-exp} of $1580$~cm$^{-1}$
than previous LDA calculations\cite{graphite-Gband-LDA}.

\begin{figure}
\includegraphics[width=0.95\columnwidth]{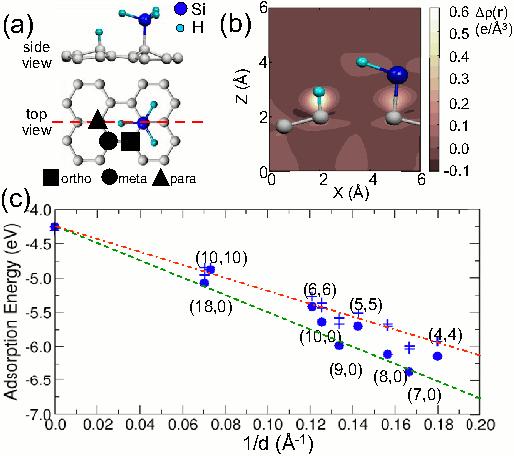}
\caption{(Color online) Bonding of SiH$_3$ and H to graphitic
carbon. (a) Equilibrium adsorption geometry on graphene in side
view (top panel) and top view (bottom panel). (b) Charge density
difference ${\Delta}{\rho}({\bf r})=\rho{\rm
(SiH_3+H/graphene)}-\rho{\rm (SiH_3)}-\rho{\rm (H)}-\rho{\rm
(graphene)}$, displayed in the plane indicated by the dashed line
in (a). (c) Adsorption energy $E_{ad}$(SiH$_3$+H) on carbon
nanotubes as a function of the inverse tube diameter $1/d$.
\label{Fig1} }
\end{figure}

We first considered co-adsorption of SiH$_3$ and H on graphene, a
good model for a wide-diameter carbon nanotube. Our calculations
suggest that both silyl and hydrogen preferentially adsorb on-top
of carbon atoms, where they interact with the initially unfilled
C$2p_z$ orbital. We found a weak mutual attraction between silyl
and hydrogen, mediated by the graphene substrate. The different
adsorption geometries in neighboring adsorption sites are
indicated schematically in Fig.~\ref{Fig1}(a). For an isolated
silyl-hydrogen pair, we distinguished the {\em ortho} geometry
with the adsorbates adjacent from the {\em para} arrangement with
the adsorbates on opposite corners of a single C$_6$ ring. In the
{\em meta} arrangement, the adsorbates are neither adjacent nor
opposite to each other on a single C$_6$ ring. Defining the
adsorption energy $E_{ad}$(SiH$_3$+H) as the energy to adsorb
SiH$_3$ and H from the vacuum, we find $E_{ad}=-4.25$~eV for {\em
ortho}, $-3.57$~eV for {\em meta}, and $-4.31$~eV for {\em para}
geometry. In view of the fact that the adsorption energy of a
hydrogen atom on graphene is close to $-1.4$~eV\cite{DT193}, we
conclude that the bond strength between the silyl radical and
graphene is unusually large, exceeding $2$~eV.

We find that the SiH$_4$(ad)$\rightarrow$SiH$_3$(ad)+H(ad)
dissociation on graphene, associated with a local pyramidalization
of the initially planar graphene near the adsorption site, occurs
with an energy barrier of ${\approx}1.5$~eV in the {\em para}
arrangement. The pyramidalization seen in the optimum adsorption
geometry, shown in side view in Fig.~\ref{Fig1}(a), indicates a
local change of the bonding character from $sp^2$ to $sp^3$, so
that initially inert planar graphene is transformed into a
reactive substrate capable of forming covalent bonds with
adsorbates. The nature of the adsorption bond can be best seen by
inspecting the charge density difference ${\Delta}{\rho}$,
representing the charge redistribution in the system upon
adsorption. The spatial distribution of ${\Delta}{\rho}({\bf r})$
for silyl and hydrogen adsorbed on graphene in the {\em para}
arrangement is shown in Fig.~\ref{Fig1}(b) in a plane normal to
the substrate. We observe no charge flow between adsorbate and
substrate, but rather a moderate charge accumulation in the Si-C
and H-C bond regions, indicating the formation of covalent bonds.

In contrast to graphene, the finite surface curvature of pristine
nanotubes introduces partial $sp^3$ character, making these
systems more reactive. This is best illustrated by plotting the
adsorption energy $E_{ad}$(SiH$_3$+H) as a function of the
nanotube diameter in Fig.~\ref{Fig1}(c) for one SiH$_3$ and one H
per supercell\cite{CNT-supercells}. As the $sp^3$ character
increases with decreasing nanotube diameter, there is a decreasing
energy cost to distort the substrate in order to optimize the
adsorption bond, thus making adsorption more exothermic. Also at
high coverages, we expect the adsorption bonds to be stronger on
the average than at low coverages. In contrast to the first
adsorbates in the supercell that introduce local substrate
distortion, bonding of additional adsorbates benefits
energetically from the previously enhanced $sp^3$ bonding
character. Depending on the adsorption arrangement within the
supercell, we have found a strengthening of the average adsorption
bond by up to $0.4$~eV when increasing the coverage from one to
three SiH$_3$ and H units per supercell.

Our results for the most stable adsorption arrangement on an
$(n,m)$ nanotube are given by solid data points, and less stable
arrangements by the crosses in Fig.~\ref{Fig1}(c). With respect to
our graphene results, the adsorption energy differences between
the {\em ortho}, {\em meta} and {\em para} arrangements in $(n,m)$
nanotubes may change by up to a few tenths of an eV per supercell,
thus changing the most favored adsorption arrangement. At nonzero
coverages, we find that silyls and hydrogens prefer to form lines
in axial direction, thus minimizing the strain energy in the
substrate. Among the systems investigated here, we found silyl and
hydrogen to bond more strongly to zigzag than to armchair
nanotubes due to differences in the pyramidalization strain.

\begin{figure}
\includegraphics[width=0.9\columnwidth]{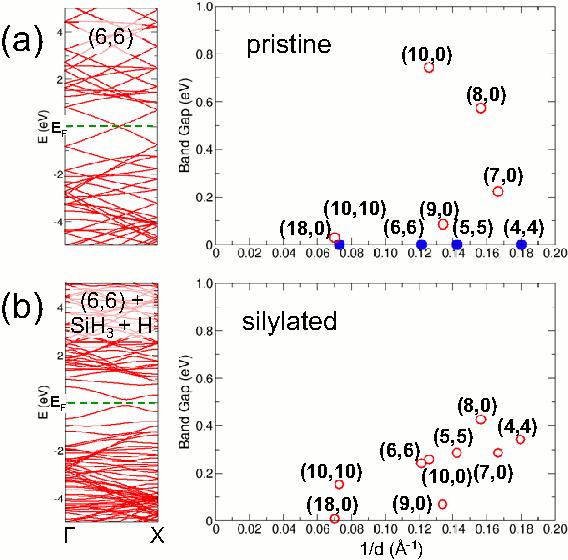}
\caption{(Color online) Electronic structure of (a) pristine and
(b) silylated carbon nanotubes. The left panels display the effect
of silylation on the band structure of the $(6,6)$ nanotube. The
right panels show the fundamental band gap $E_g$ in nanotubes as a
function of the inverse tube diameter $1/d$. Data for metallic
nanotubes are given by solid symbols, those for semiconducting
nanotubes by open symbols. \label{Fig2} }
\end{figure}

Having established that silyl radicals and hydrogen form strong
covalent bonds with nanotubes, we next investigate the effect of
chemisorption on the electronic structure of carbon nanotubes. As
mentioned earlier, the fundamental band gap of pristine carbon
nanotubes varies strongly with the chiral index
$(n,m)$\cite{{Mintmire92},{RSaito92},{Oshiyama92}}. In density
functional calculations, the band gap value is known to be
underestimated significantly. Consequently, the range of band gap
values in nanotubes should be even larger than what is suggested
by the DFT results for pristine nanotubes in the right panel of
Fig.~\ref{Fig2}(a). In a sample with a random distribution of
chiral indices $(n,m)$, we expect one third of nanotubes to be
metallic\cite{{Mintmire92},{RSaito92},{Oshiyama92}}. For better
visual impact, metallic nanotubes are represented by solid data
points and semiconducting nanotubes by open data points in the
right panel of Fig.~\ref{Fig2}(a). The band structure $E(k)$ of
the pristine $(6,6)$ armchair carbon nanotube is shown in the left
panel of Fig.~\ref{Fig2}(a) as a representative for other metallic
nanotubes.

The effect of silylation on the electronic structure of carbon
nanotubes is shown in Fig.~\ref{Fig2}(b). Comparing the calculated
band gaps in the silylated and pristine nanotubes in the right
panels of Fig.~\ref{Fig2}, we conclude that silylation has
converted all nanotubes to semiconductors and narrowed down the
range of their fundamental band gaps. Details of
silylation-related changes in the electronic structure can be best
seen when comparing the electronic structure of the pristine and
the silylated $(6,6)$ nanotube in the left panels of
Fig.~\ref{Fig2}. Whereas the metallic character of the pristine
system stems from the crossing of two linear bands at the Fermi
level, enhancing the $sp^3$ character by silylation opens up a
direct band gap in this system. Only in systems, where the number
of H and SiH$_3$ adsorbates per unit cell is not the same, a
non-bonding low-dispersion band appears in the band gap near
$E_F$. This impurity band is spin polarized and localized near the
adsorbates.

Obviously, the band gap of a given nanotube will change with
changing $sp^3$ character\cite{Yang2000}, which depends on the
adsorption geometry and the adsorbate coverage. Our band gap
results in the right panel of Fig.~\ref{Fig2}(b) are for a low
coverage of one SiH$_3$ and H per supercell\cite{CNT-supercells}
of the pristine nanotube. Since the $sp^3$ character should
increase with rising silyl coverage, and since DFT underestimates
the fundamental band gap, our results in the right panel of
Fig.~\ref{Fig2}(b) are to be considered a lower bound on band gaps
in silylated nanotubes.


Successful silylation of nanotubes can be verified experimentally
by inspecting the vibrational spectra as a specific signature of
silyl-nanotube bonds. Reported Raman spectra of silylated carbon
nanotubes\cite{SSWong06} suggest that the dominant effect of
silylation is the modification of mode intensities. No frequency
shifts specifically related to silylation have been observed,
suggesting that the effect of silylation may be comparable to that
of bundling\cite{SSWong06}. In the following, we determine the
vibrational spectra of pristine and silylated nanotubes in order
to identify a spectroscopic signature of the silyl-nanotube bond.

We have studied the vibrational spectra of $(6,6)$ and $(10,0)$
carbon nanotubes containing up to three silyls and hydrogens per
supercell\cite{CNT-supercells}. Upon silylation, we first expect
the emergence of new vibrational modes in the nanotube, associated
with Si-C and H-C stretch motion. Silylation furthermore modifies
the vibrational modes of the pristine nanotube due to
adsorbate-related deformations and local stress, as well as the
attachment of heavy SiH$_3$ ligands.

\begin{figure}
\includegraphics[width=1.0\columnwidth]{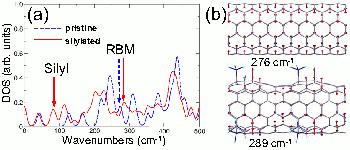}
\caption{(Color online) Effect of silylation on the vibrations of
the $(6,6)$ carbon nanotube. (a) Low-frequency range of the
vibrational spectrum. (b) Schematic of the radial breathing mode
(RBM) in the pristine and silylated nanotube. \label{Fig3} }
\end{figure}

Comparing the vibrational spectra of $(6,6)$ and $(10,0)$
nanotubes, we find the general trends for silylation-related
changes to be chirality independent, and frequency shifts to be
generally less than $20$~cm$^{-1}$. The vibrational spectra of a
pristine and a silylated $(6,6)$ carbon nanotube, with three
silyls and hydrogens in the {\em ortho} arrangement forming a ring
within the $9.9$~{\AA} long supercell\cite{CNT-supercells}, are
compared to each other in Fig.~\ref{Fig3}(a). In the low-frequency
part of the spectrum, silylated nanotubes exhibit a mode at
${\approx}80$~cm$^{-1}$, caused by the wagging motion of the
chemisorbed radicals. Other modes induced by silylation include
the high-frequency H-Si stretch in SiH$_3$ at $2000$~cm$^{-1}$ and
the H-C stretch of the adsorbed hydrogens at $2800$~cm$^{-1}$.

A vibration mode unique to carbon nanotubes is the $A_{1g}$ radial
breathing mode (RBM), shown schematically in the top panel of
Fig.~\ref{Fig3}(b). As seen in Fig.~\ref{Fig3}(a), this mode
occurs at $276$~cm$^{-1}$ in the pristine $(6,6)$ nanotube. Even
though the RBM mode is not well defined in partly silylated
nanotubes, a closely related mode, shown in the bottom panel of
Fig.~\ref{Fig3}(b), occurs at the higher frequency of
$289$~cm$^{-1}$ in the silylated nanotube. Similar RBM up-shifts
have been reported in pristine tubes and traced back to
hydrostatic
pressure\cite{{CNTRamanshift05},{CNTRamanshift-hydro03}} or stress
induced by inter-tube interaction\cite{CNTRamanshift00}. We find
an up-shift of RBM-like modes at lower coverages, when the axial
distance between adsorbed silyl radicals is large, and a
down-shift at high coverages, when most carbon atoms are
covalently bonded to a silyl. Much of this trend can be understood
and reproduced by considering carbon atoms connected to silyls as
virtual `heavy' $^{43}$C `isotopes' within the pristine nanotube.
In the low-coverage regime considered in Fig.~\ref{Fig3}(b), the
heavy carbons move radially out of phase with the other carbons,
causing a frequency up-shift. At high coverages, all carbons are
subject to the usual radial breathing motion with the frequency
down-shifted by the isotope effect.

The high-frequency tangential mode, also known as the G band, is
down-shifted in the silylated nanotube to $1619$~cm$^{-1}$ from
its initial value of $1638$~cm$^{-1}$ in the pristine $(6,6)$
nanotube. This frequency shift is almost independent of the silyl
coverage.

In conclusion, we used {\em ab initio} density functional
calculations to study the chemical functionalization of
single-wall carbon nanotubes and graphene monolayers by silyl
radicals and hydrogen. We found that silyl radicals form strong
covalent bonds with graphene and nanotube walls, causing local
structural relaxations that enhance the $sp^3$ character of these
graphitic nanostructures. With the help of silane chemistry,
nanotube walls may form strong, well-defined bonds to a
surrounding matrix, facilitating the transition to carbon-silicon
hybrid electronics. We found that silylation transforms all carbon
nanotubes into semiconductors, independent of their chirality. Our
calculated vibrational spectra suggest that specific frequency
shifts can be used as a signature of successful silylation.


We acknowledge useful discussions with Yung-Doug Suh and Hee-Cheul
Choi about silylation of carbon nanotubes and the assistance of
Shinya Okano with numerical calculations. This work was funded by
the National Science Foundation under NSF-NSEC grant 425826 and
NSF-NIRT grant ECS-0506309. Computational resources have been
provided by the Michigan State University High Performance
Computing Center.


\begin{thebibliography}{22}
\expandafter\ifx\csname
natexlab\endcsname\relax\def\natexlab#1{#1}\fi
\expandafter\ifx\csname bibnamefont\endcsname\relax
  \def\bibnamefont#1{#1}\fi
\expandafter\ifx\csname bibfnamefont\endcsname\relax
  \def\bibfnamefont#1{#1}\fi
\expandafter\ifx\csname citenamefont\endcsname\relax
  \def\citenamefont#1{#1}\fi
\expandafter\ifx\csname url\endcsname\relax
  \def\url#1{\texttt{#1}}\fi
\expandafter\ifx\csname
urlprefix\endcsname\relax\def\urlprefix{URL }\fi
\providecommand{\bibinfo}[2]{#2}
\providecommand{\eprint}[2][]{\url{#2}}

\bibitem[{\citenamefont{Jorio et~al.}(2008)\citenamefont{Jorio, Dresselhaus,
  and Dresselhaus}}]{TAP111}
\bibinfo{author}{\bibfnamefont{A.}~\bibnamefont{Jorio}},
  \bibinfo{author}{\bibfnamefont{M.}~\bibnamefont{Dresselhaus}},
  \bibnamefont{and}
  \bibinfo{author}{\bibfnamefont{G.}~\bibnamefont{Dresselhaus}},
  \emph{\bibinfo{title}{Carbon Nanotubes: Advanced Topics in the Synthesis,
  Structure, Properties and Applications}}, vol. \bibinfo{volume}{111} of
  \emph{\bibinfo{series}{Topics in Applied Physics}}
  (\bibinfo{publisher}{Springer}, \bibinfo{address}{Berlin},
  \bibinfo{year}{2008}).

\bibitem[{\citenamefont{Nemec et~al.}(2006)\citenamefont{Nemec, Tomanek, and
  Cuniberti}}]{DT178}
\bibinfo{author}{\bibfnamefont{N.}~\bibnamefont{Nemec}},
  \bibinfo{author}{\bibfnamefont{D.}~\bibnamefont{Tomanek}}, \bibnamefont{and}
  \bibinfo{author}{\bibfnamefont{G.}~\bibnamefont{Cuniberti}},
  \bibinfo{journal}{Phys. Rev. Lett.} \textbf{\bibinfo{volume}{96}},
  \bibinfo{eid}{076802} (\bibinfo{year}{2006}).

\bibitem[{\citenamefont{Mintmire et~al.}(1992)\citenamefont{Mintmire, Dunlap,
  and White}}]{Mintmire92}
\bibinfo{author}{\bibfnamefont{J.~W.} \bibnamefont{Mintmire}},
  \bibinfo{author}{\bibfnamefont{B.~I.} \bibnamefont{Dunlap}},
  \bibnamefont{and} \bibinfo{author}{\bibfnamefont{C.~T.} \bibnamefont{White}},
  \bibinfo{journal}{Phys. Rev. Lett.} \textbf{\bibinfo{volume}{68}},
  \bibinfo{pages}{631} (\bibinfo{year}{1992}).

\bibitem[{\citenamefont{Saito et~al.}(1992)\citenamefont{Saito, Fujita,
  Dresselhaus, and Dresselhaus}}]{RSaito92}
\bibinfo{author}{\bibfnamefont{R.}~\bibnamefont{Saito}},
  \bibinfo{author}{\bibfnamefont{M.}~\bibnamefont{Fujita}},
  \bibinfo{author}{\bibfnamefont{G.}~\bibnamefont{Dresselhaus}},
  \bibnamefont{and} \bibinfo{author}{\bibfnamefont{M.~S.}
  \bibnamefont{Dresselhaus}}, \bibinfo{journal}{Appl. Phys. Lett.}
  \textbf{\bibinfo{volume}{60}}, \bibinfo{pages}{2204} (\bibinfo{year}{1992}).

\bibitem[{\citenamefont{Hamada et~al.}(1992)\citenamefont{Hamada, Sawada, and
  Oshiyama}}]{Oshiyama92}
\bibinfo{author}{\bibfnamefont{N.}~\bibnamefont{Hamada}},
  \bibinfo{author}{\bibfnamefont{S.~I.} \bibnamefont{Sawada}},
  \bibnamefont{and} \bibinfo{author}{\bibfnamefont{A.}~\bibnamefont{Oshiyama}},
  \bibinfo{journal}{Phys. Rev. Lett.} \textbf{\bibinfo{volume}{68}},
  \bibinfo{pages}{1579} (\bibinfo{year}{1992}).

\bibitem[{\citenamefont{Hemraj-Benny and Wong}(2006)}]{SSWong06}
\bibinfo{author}{\bibfnamefont{T.}~\bibnamefont{Hemraj-Benny}}
  \bibnamefont{and} \bibinfo{author}{\bibfnamefont{S.~S.} \bibnamefont{Wong}},
  \bibinfo{journal}{Chem. Mater.} \textbf{\bibinfo{volume}{18}},
  \bibinfo{pages}{4827} (\bibinfo{year}{2006}).

\bibitem[{\citenamefont{Weissenbach and Mack}(2005)}]{silane-bonding}
\bibinfo{author}{\bibfnamefont{K.}~\bibnamefont{Weissenbach}} \bibnamefont{and}
  \bibinfo{author}{\bibfnamefont{H.}~\bibnamefont{Mack}}, in
  \emph{\bibinfo{booktitle}{Functional Fillers for Plastics}}, edited by
  \bibinfo{editor}{\bibfnamefont{M.}~\bibnamefont{Xanthos}}
  (\bibinfo{publisher}{WILEY-VCH}, \bibinfo{address}{Weinheim},
  \bibinfo{year}{2005}), \bibinfo{type}{chapter}~\bibinfo{chapter}{4}, pp.
  \bibinfo{pages}{57--84}.

\bibitem[{\citenamefont{Pierce}(1968)}]{silane-book}
\bibinfo{author}{\bibfnamefont{A.~E.} \bibnamefont{Pierce}},
  \emph{\bibinfo{title}{Silylation of organic compounds}}
  (\bibinfo{publisher}{Pierce Chemical Company}, \bibinfo{address}{Rockford,
  Illinois}, \bibinfo{year}{1968}).

\bibitem[{\citenamefont{Perdew and Zunger}(1981)}]{Perdew81}
\bibinfo{author}{\bibfnamefont{J.~P.} \bibnamefont{Perdew}} \bibnamefont{and}
  \bibinfo{author}{\bibfnamefont{A.}~\bibnamefont{Zunger}},
  \bibinfo{journal}{Phys. Rev. B} \textbf{\bibinfo{volume}{23}},
  \bibinfo{pages}{5048} (\bibinfo{year}{1981}).

\bibitem[{\citenamefont{Soler et~al.}(2002)\citenamefont{Soler, Artacho, Gale,
  Garc\'{i}a, Junquera, Ordej\'{o}n, and
  S\'{a}nchez-Portal}}]{soler-tsmfaioms2002}
\bibinfo{author}{\bibfnamefont{J.~M.} \bibnamefont{Soler}},
  \bibinfo{author}{\bibfnamefont{E.}~\bibnamefont{Artacho}},
  \bibinfo{author}{\bibfnamefont{J.~D.} \bibnamefont{Gale}},
  \bibinfo{author}{\bibfnamefont{A.}~\bibnamefont{Garc\'{i}a}},
  \bibinfo{author}{\bibfnamefont{J.}~\bibnamefont{Junquera}},
  \bibinfo{author}{\bibfnamefont{P.}~\bibnamefont{Ordej\'{o}n}},
  \bibnamefont{and}
  \bibinfo{author}{\bibfnamefont{D.}~\bibnamefont{S\'{a}nchez-Portal}},
  \bibinfo{journal}{J. Phys: Condens. Matter} \textbf{\bibinfo{volume}{14}},
  \bibinfo{pages}{2745} (\bibinfo{year}{2002}).

\bibitem[{\citenamefont{Troullier and Martins}(1991)}]{Troullier91}
\bibinfo{author}{\bibfnamefont{N.}~\bibnamefont{Troullier}} \bibnamefont{and}
  \bibinfo{author}{\bibfnamefont{J.~L.} \bibnamefont{Martins}},
  \bibinfo{journal}{Phys. Rev. B} \textbf{\bibinfo{volume}{43}},
  \bibinfo{pages}{1993} (\bibinfo{year}{1991}).

\bibitem[{\citenamefont{Kleinman and Bylander}(1982)}]{Kleinman82}
\bibinfo{author}{\bibfnamefont{L.}~\bibnamefont{Kleinman}} \bibnamefont{and}
  \bibinfo{author}{\bibfnamefont{D.~M.} \bibnamefont{Bylander}},
  \bibinfo{journal}{Phys. Rev. Lett.} \textbf{\bibinfo{volume}{48}},
  \bibinfo{pages}{1425} (\bibinfo{year}{1982}).

\bibitem[{CNT()}]{CNT-supercells}
\bibinfo{note}{We used supercells of comparable length, containing 4 primitive
  unit cells for armchair nanotubes and 2 primitive unit cells for zigzag
  nanotubes.}

\bibitem[{\citenamefont{Artacho et~al.}(1999)\citenamefont{Artacho,
  S\'{a}nchez-Portal, Ordej\'{o}n, Garc\'{\i}a, and Soler}}]{SIESTA_PAO}
\bibinfo{author}{\bibfnamefont{E.}~\bibnamefont{Artacho}},
  \bibinfo{author}{\bibfnamefont{D.}~\bibnamefont{S\'{a}nchez-Portal}},
  \bibinfo{author}{\bibfnamefont{P.}~\bibnamefont{Ordej\'{o}n}},
  \bibinfo{author}{\bibfnamefont{A.}~\bibnamefont{Garc\'{\i}a}},
  \bibnamefont{and} \bibinfo{author}{\bibfnamefont{J.~M.} \bibnamefont{Soler}},
  \bibinfo{journal}{Phys. Stat. Sol.} \textbf{\bibinfo{volume}{215}},
  \bibinfo{pages}{809} (\bibinfo{year}{1999}).

\bibitem[{gra()}]{graphite-Gband-LDAGGA}
\bibinfo{note}{Whereas LDA provides a more consistent picture of bonding to
  $sp^2$ carbons, the GGA force field sometimes achieves better agreement with
  experimental results. Our GGA value of $1582$~cm$^{-1}$ for the G-band
  frequency is in perfect agreement with the experimental value.}

\bibitem[{\citenamefont{Ferrari et~al.}(2006)\citenamefont{Ferrari, Meyer,
  Scardaci, Casiraghi, Lazzeri, Mauri, Piscanec, Jiang, Novoselov, Roth
  et~al.}}]{graphite-Gband-exp}
\bibinfo{author}{\bibfnamefont{A.~C.} \bibnamefont{Ferrari}},
  \bibinfo{author}{\bibfnamefont{J.~C.} \bibnamefont{Meyer}},
  \bibinfo{author}{\bibfnamefont{V.}~\bibnamefont{Scardaci}},
  \bibinfo{author}{\bibfnamefont{C.}~\bibnamefont{Casiraghi}},
  \bibinfo{author}{\bibfnamefont{M.}~\bibnamefont{Lazzeri}},
  \bibinfo{author}{\bibfnamefont{F.}~\bibnamefont{Mauri}},
  \bibinfo{author}{\bibfnamefont{S.}~\bibnamefont{Piscanec}},
  \bibinfo{author}{\bibfnamefont{D.}~\bibnamefont{Jiang}},
  \bibinfo{author}{\bibfnamefont{K.~S.} \bibnamefont{Novoselov}},
  \bibinfo{author}{\bibfnamefont{S.}~\bibnamefont{Roth}}, \bibnamefont{et~al.},
  \bibinfo{journal}{Phys. Rev. Lett.} \textbf{\bibinfo{volume}{97}},
  \bibinfo{eid}{187401} (\bibinfo{year}{2006}).

\bibitem[{\citenamefont{Reich et~al.}(2001)\citenamefont{Reich, Thomsen, and
  Ordej\'on}}]{graphite-Gband-LDA}
\bibinfo{author}{\bibfnamefont{S.}~\bibnamefont{Reich}},
  \bibinfo{author}{\bibfnamefont{C.}~\bibnamefont{Thomsen}}, \bibnamefont{and}
  \bibinfo{author}{\bibfnamefont{P.}~\bibnamefont{Ordej\'on}},
  \bibinfo{journal}{Phys. Rev. B} \textbf{\bibinfo{volume}{64}},
  \bibinfo{pages}{195416} (\bibinfo{year}{2001}).

\bibitem[{\citenamefont{Miller et~al.}(2008)\citenamefont{Miller, Kintigh, Kim,
  Weck, Berber, and Tomanek}}]{DT193}
\bibinfo{author}{\bibfnamefont{G.}~\bibnamefont{Miller}},
  \bibinfo{author}{\bibfnamefont{J.}~\bibnamefont{Kintigh}},
  \bibinfo{author}{\bibfnamefont{E.}~\bibnamefont{Kim}},
  \bibinfo{author}{\bibfnamefont{P.}~\bibnamefont{Weck}},
  \bibinfo{author}{\bibfnamefont{S.}~\bibnamefont{Berber}}, \bibnamefont{and}
  \bibinfo{author}{\bibfnamefont{D.}~\bibnamefont{Tomanek}},
  \bibinfo{journal}{J. Am. Chem. Soc.} \textbf{\bibinfo{volume}{130}},
  \bibinfo{pages}{2296} (\bibinfo{year}{2008}).

\bibitem[{\citenamefont{Yang and Han}(2000)}]{Yang2000}
\bibinfo{author}{\bibfnamefont{L.}~\bibnamefont{Yang}} \bibnamefont{and}
  \bibinfo{author}{\bibfnamefont{J.}~\bibnamefont{Han}},
  \bibinfo{journal}{Phys. Rev. Lett.} \textbf{\bibinfo{volume}{85}},
  \bibinfo{pages}{154} (\bibinfo{year}{2000}).

\bibitem[{\citenamefont{Lefrant et~al.}(2005)\citenamefont{Lefrant, Baltog, and
  Baibarac}}]{CNTRamanshift05}
\bibinfo{author}{\bibfnamefont{S.}~\bibnamefont{Lefrant}},
  \bibinfo{author}{\bibfnamefont{I.}~\bibnamefont{Baltog}}, \bibnamefont{and}
  \bibinfo{author}{\bibfnamefont{M.}~\bibnamefont{Baibarac}},
  \bibinfo{journal}{J. Raman Spectroscopy} \textbf{\bibinfo{volume}{36}},
  \bibinfo{pages}{676} (\bibinfo{year}{2005}).

\bibitem[{\citenamefont{Sandler et~al.}(2003)\citenamefont{Sandler, Shaffer,
  Windle, Halsall, Montes-Mor\'an, Cooper, and Young}}]{CNTRamanshift-hydro03}
\bibinfo{author}{\bibfnamefont{J.}~\bibnamefont{Sandler}},
  \bibinfo{author}{\bibfnamefont{M.~S.~P.} \bibnamefont{Shaffer}},
  \bibinfo{author}{\bibfnamefont{A.~H.} \bibnamefont{Windle}},
  \bibinfo{author}{\bibfnamefont{M.~P.} \bibnamefont{Halsall}},
  \bibinfo{author}{\bibfnamefont{M.~A.} \bibnamefont{Montes-Mor\'an}},
  \bibinfo{author}{\bibfnamefont{C.~A.} \bibnamefont{Cooper}},
  \bibnamefont{and} \bibinfo{author}{\bibfnamefont{R.~J.} \bibnamefont{Young}},
  \bibinfo{journal}{Phys. Rev. B} \textbf{\bibinfo{volume}{67}},
  \bibinfo{pages}{035417} (\bibinfo{year}{2003}).

\bibitem[{\citenamefont{Dresselhaus and Eklund}(2000)}]{CNTRamanshift00}
\bibinfo{author}{\bibfnamefont{M.~S.} \bibnamefont{Dresselhaus}}
  \bibnamefont{and} \bibinfo{author}{\bibfnamefont{P.~C.}
  \bibnamefont{Eklund}}, \bibinfo{journal}{Adv. Phys.}
  \textbf{\bibinfo{volume}{49}}, \bibinfo{pages}{705} (\bibinfo{year}{2000}).

\end{thebibliography}

\end{document}